\documentstyle[12pt,epsfig]{article}
\newcommand{\vc}[1]{\vec{#1}} 
\newcommand{\nuc}[2]{$^{#1}${#2}} 
\newcommand{\e}{{\mathrm{e}}} 
\renewcommand{\d}{{\mathrm{d}}} 
\begin{document} 
\small 
\begin{center} 
{\bf 
	   The Two-Body Momentum Distribution 
           in Finite Nuclei 
	   }\\[5mm]   
{P. Papakonstantinou}$^a$, 
{E. Mavrommatis}$^a$ and  
{T.S. Kosmas}$^b$\\[5mm]  
{\em $^a${Physics Department, Division of Nuclear and Particle 
Physics,
University of Athens, GR-15771 Athens, Greece}\\ 
$^b${Physics Department,Theoretical Physics Division, 
University of Ioannina, GR-45110 Ioannina, Greece} 
  } 
\end{center}

%

\noindent 
\bf Abstract   
\rm 

\noindent 
The two-body momentum distribution 
$\eta_2(\vc{p}_1,\vc{p}_2)$ 
of nuclei is studied. 
First, a compact analytical expression is derived 
for $Z=N$, $\ell$-closed nuclei, 
within the context 
of the independent particle shell model. 
Application to the light closed-shell nucleus
\nuc{16}{O} 
is included and discussed. 
Next, the effect of dynamical, 
short-range correlations is investigated
in the case of the light nucleus \nuc{4}{He}, 
by including Jastrow-type correlations in 
the formalism. 
The effect is significant for 
large values of $p_1$ and $p_2$ 
and for angles between the vectors 
$\vc{p}_1$ and $\vc{p}_2$ close to 
$\gamma =$180$^{\circ}$ and 0$^{\circ}$. 
%
\section{Introduction} 
The problem of 
short-range nucleon-nucleon correlations (SRC) in nuclei 
is still an open field in nuclear physics 
\cite{AHP93}. 
An indirect evidence for SRC 
was obtained by the observation of a depletion of valence shells 
in $(e,e'p)$ reactions \cite{La93,PS97}. 
It is also 
well known that SRC 
modify the nucleon momentum distribution \cite{AHP88,MP00}, 
i.e. the Fourier transform of the 
one-body density matrix $\rho_1 (\vc{r}_1,\vc{r}_{1'})$ 
in the variable 
$\vc{r}_1-\vc{r}_{1'}$, 
by introducing significant contributions 
at values of momentum beyond the Fermi momentum. 
Experimental information from 
inclusive $(e,e')$ and exclusive $(e,e'p)$ reactions 
established the existence of a high-momentum component 
in the momentum distribution. 

Beyond the one-body density matrix 
and momentum distribution, 
rich information on the nuclear ground state and 
nucleon-nucleon correlations 
is contained in the two-body density matrix (2DM) 
$\rho_2 (\vc{r}_1,\vc{r}_2;\vc{r}_{1'},\vc{r}_{2'})$ 
and its various Fourier transforms in momentum space. 
Experimentally, 
exclusive two-nucleon knockout reactions 
can provide information on the 
relative behaviour of nucleon pairs embedded 
in the nuclear medium 
and give more direct insight into SRC 
\cite{BF90,Ja95,Ro00,St00,Ne98,Ba91,Wa00,Ac99}. 
A very important role is played by electron-nucleus 
scattering experiments 
with the availability of high-energy 100\% 
duty cycle electron beams which made possible to carry out 
double coincidence $(e,e'NN)$ experiments. 
Past, present and near future experiments at 
NIKHEF, MAMI and Jefferson Lab 
use \nuc{3}{He}, \nuc{12}{C} and \nuc{16}{O} as targets 
\cite{Ro00,St00,Ne98,Ba91}. 

The analysis of $(e,e'2N)$ experiments is a 
complicated task 
\cite{BGP96,Ge96,Gi98,Ry97}. 
Among the most important processes of longitudinal character 
contributing to electromagnetically induced 
two-nucleon knockout is the scattering off a 
correlated nucleon pair followed 
by two-nucleon knockout 
during which each of the two nucleons interacts 
with the residual nucleus 
(final-state interactions - FSI). 
The minimal starting point in the analysis 
is the plane-wave impulse approximation (PWIA), 
in which the FSI are neglected and 
the two-nucleon knock-out cross section 
can be expressed in terms of the two-hole spectral function
$S(\vc{p}_1,\vc{p}_2;E)$. 
The latter gives the joint probability of removing from the 
target two nucleons with momenta $\vc{p}_1$ and $\vc{p}_2$ 
leaving the residual nucleus with energy $E$ 
with respect to the target nucleus ground state. 
Under certain conditions it seems possible to extract information 
on the 2DM and its Fourier transforms 
and on nucleon-nucleon correlations. 
Taking the above into account,
the development of a simple model of the ground-state 
2DM in order to clarify the two-body characteristics 
of the nuclear ground state 
and the extent to which it is affected by the SRC 
appears useful. 

Here we focus on the 
two-body momentum distribution (2bMD)
$\eta_2(\vc{p}_1,\vc{p}_2)$, which is 
the Fourier transform of the 2DM 
$\rho_2(\vc{r}_1,\vc{r}_2;\vc{r}_{1'},\vc{r}_{2'})$ 
in the variables 
$\vc{r}_1-\vc{r}_{1'}$ 
and 
$\vc{r}_2-\vc{r}_{2'}$. 
Another Fourier transform of the 
half-diagonal two-body density matrix 
$\rho_{2h}(\vc{r}_1,\vc{r}_2,\vc{r}_{1'}) 
 = \rho_2 (\vc{r}_1,\vc{r}_2;\vc{r}_{1'},\vc{r}_2)$ \cite{PMC01}, 
namely the generalized momentum distribution $\eta (\vc{p},\vc{Q})$, 
has been studied previously in the case of 
infinite nuclear matter \cite{MPC95} and finite nuclei \cite{PMK00}. 
The generalized momentum distribution 
plays an important role 
in the study of the FSI. 
The 2bMD is straighforwardly connected to the 
two-nucleon spectral function 
$S(\vc{p}_1,\vc{p}_2;E)$ 
via an integration with respect to the energy 
(see Eq.(\ref{Espectral}) below).
The two-nucleon spectral function 
in the case of infinite nuclear matter 
has been studied using correlated basis function 
perturbation theory including central 
and tensor correlations \cite{BF00}. 
In the case of \nuc{16}{O} 
it has been evaluated 
including short-range correlations 
through defect functions calculated 
via the Bethe-Goldstone equation or 
correlation functions derived from 
variational calculations and long-range 
correlations using dressed RPA \cite{Ge96}.

In this work, first  an analytical expression is derived 
for the 2bMD of $Z=N$, $\ell$-closed nuclei 
in the independent-particle shell model 
with harmonic oscillator wave-functions 
and applied to the nucleus \nuc{16}{O}. 
Our method 
is 
an extension of that 
developed in Refs.  
\cite{PMK00,KV90,KV92,Pa95e} 
for the study of the nuclear form factor,
the nuclear charge, matter and momentum distributions, 
the one-body density matrix 
and the generalized momentum distribution 
in closed shell nuclei. 
This calculation is expected to reproduce the main features 
of the 2bMD, 
stemming from the finite nuclear size and Fermi statistics, 
being reliable at certain kinematical domains. 
By introducing the quantity 
$\eta_2(\vc{p}_1,\vc{p}_2)/\eta (\vc{p}_1)\eta(\vc{p}_2)$, 
where $\eta (\vc{p})$ is the nucleon momentum distribution, 
we pinpoint the effects of 
the finite size and of 
the statistical correlations. 

Next, the effect of the SRC on the 2bMD of \nuc{4}{He} 
is studied by including Jastrow-type correlations 
in our calculation, 
via the lowest term of a cluster expansion. 
This so-called low order approximation (LOA) 
\cite{GGR71} 
has been exploited 
for the one-body density matrix 
and the two-body density matrix 
by 
Bohigas and Stringari \cite{BS80} 
and Dal R\`{i}, Stringari and Bohigas \cite{DSB82}
and 
has been 
widely used to study 
single-particle nuclear properties, such as 
the (point-nucleon or charge) density \cite{SAD93b,MM99}, 
the elastic or charge form factor \cite{BS80,DSB82,MM99}, 
the one-body density matrix in 
coordinate \cite{SAD93b,MM00}  
or momentum space \cite{SAD93b}  
and the nucleon momentum distribution 
\cite{BS80,DSB82,SAD93b,MM00},   
as well as two-body nuclear quantities, 
namely 
the two-body density matrix in coordinate space \cite{DKA00}, 
the static structure function $S(Q)$ \cite{DSB82}, 
the relative pair density distribution $\rho_2(r_{12})$ 
\cite{DSB82,DKA00}, 
the center-of-mass pair density distribution $\rho_2(R_{CM})$ 
\cite{DKA00},
the two-body 
relative momentum distribution 
$\eta_2(\vc{q}_{\mathrm rel})$ \cite{DKA00,OS95}, 
the two-body 
center-of-mass momentum distribution 
$\eta_2(\vc{P}_{\mathrm CM})$ \cite{DKA00,OS95} 
and the combined two-body 
center-of-mass and relative 
momentum distribution 
$\eta_2(P_{\mathrm CM},q_{\mathrm rel})$ \cite{OS95}. 
A two-gaussian central correlation function, 
$f(r)=1-c_1\exp (-r^2/\beta_1^2)+c_2\exp (-r^2/\beta_2^2)$, 
is used in this work.
We find significant deviations from the independent-particle picture 
at high momenta, especially when the vectors 
$\vc{p}_1$ and $\vc{p}_2$ are antiparallel or parallel, 
as it is clearly shown by the values
of the quantity 
$\eta_2(\vc{p}_1,\vc{p}_2)/\eta (\vc{p}_1)\eta(\vc{p}_2)$. 
There is, however, within the LOA, significant sensitivity of the results 
in some kinematical regions, mainly on the correlation function used 
to describe the SRC. 
For certain correlation functions 
one may obtain 
negative values of $\eta_2$ 
in few regions of momenta. 

The definition and properties 
of the two-body momentum distribution are presented 
in detail in Section~2. 
In Section~\ref{Smoc} 
an exact expression is derived for the 2bMD 
of $Z=N$, $\ell$-closed nuclei 
in the independent-particle shell model 
with harmonic oscillator wavefunctions 
and applied to the case of \nuc{16}{O}. 
In Section~\ref{Sdc} the effect of SRC  
on the 2bMD of \nuc{4}{He}  
is explored by including Jastrow-type correlations in our calculation 
via the LOA, 
and the results as well as the 
reliability of our approximation are discussed. 
Finally, in Section~\ref{Ssap} we give 
a summary of the results and 
hints for possible future developments. 

\section{The Two-Body Momentum Distribution} 
The two-body momentum distribution 
$\eta_2(\vc{p}_1,\vc{p}_2)$ 
gives the combined probability density 
of finding in the nucleus a nucleon with momentum $\vc{p}_1$ 
and another one with momentum $\vc{p}_2$. 
In a system of $A$ $(A\geq 2)$ identical particles, 
described by a unit-normalized state $|\Psi\rangle$, 
it can be defined as the expectation value 
\begin{equation} 
\eta_2(\vc{p}_1,\vc{p}_2) = 
  \langle\Psi | 
  \sum_{\vc{s},\vc{s'}} 
   a^{\dagger}_{\vc{p}_2,\vc{s'}}  
   a^{\dagger}_{\vc{p}_1,\vc{s}}  
   a_{\vc{p}_1,\vc{s}}  
   a_{\vc{p}_2,\vc{s'}} |\Psi\rangle . 
\end{equation} 
The 2bMD is the diagonal element of the 2DM in momentum space 
$\eta_2(\vc{p}_1,\vc{p}_2;\vc{p}_{1'},\vc{p}_{2'})$, 
\begin{equation} 
\eta_2(\vc{p}_1,\vc{p}_2) = 
\eta_2(\vc{p}_1,\vc{p}_2;\vc{p}_1,\vc{p}_2) 
\end{equation} 
and may also be defined as the Fourier transform of the 2DM 
$\rho_2(\vc{r}_1,\vc{r}_2;\vc{r}_{1'},\vc{r}_{2'})$ 
in the variables 
$\vc{r}_1-\vc{r}_{1'}$ 
and 
$\vc{r}_2-\vc{r}_{2'}$: 
\begin{equation} 
\eta_2(\vc{p}_1,\vc{p}_2) = 
\int 
\rho_2(\vc{r}_1,\vc{r}_2;\vc{r}_{1'},\vc{r}_{2'}) 
\e^{-i\vc{p}_1\cdot (\vc{r}_1-\vc{r}_{1'})}  
\e^{-i\vc{p}_2\cdot (\vc{r}_2-\vc{r}_{2'})}  
{\d^3}r_1{\d^3}r_2{\d^3}r_{1'}{\d^3}r_{2'}. 
\label{Edft} 
\end{equation} 
(By taking 
$\hbar =1$, 
the 2bMD has the dimension (length)$^6$.) 
The normalization adopted is such that 
\begin{equation} 
\int 
\eta_2(\vc{p}_1,\vc{p}_2) {\d^3}p_1{\d^3}p_2 = 
A(A-1) , 
\label{Enor} 
\end{equation} 
and the following sequential relation 
holds: 
\begin{equation} 
\int 
\eta_2(\vc{p}_1,\vc{p}_2) {\d^3}p_2 = 
(A-1)\eta (\vc{p}_1) , 
\label{Eseq} 
\end{equation} 
where $\eta (\vc{p})$ is the one-body momentum distribution, 
normalized to the number of particles 
\begin{equation} 
\int 
\eta (\vc{p}) {\d^3}p = 
A . 
\label{Enor1} 
\end{equation}
Equation~(\ref{Eseq}) connects the 2bMD to the 
generalized momentum distribution 
\cite{MPC95,PMK00} 
($\eta (\vc{p},\vc{Q}=0) = (A-1)\eta (\vc{p})$). 

The 2bMD 
$\eta_2(\vc{p}_1,\vc{p}_2)$ 
is related to the two-body  
center-of-mass momentum distribution 
$\eta_2(\vc{P}_{\mathrm CM})$ 
($\vc{P}_{\mathrm CM}=\vc{p}_1+\vc{p}_2$ 
is the two-nucleon center-of-mass momentum),  
and to the two-body relative momentum distribution 
$\eta_2(\vc{q}_{\mathrm rel})$ 
($\vc{q}_{\mathrm rel}=
\frac{1}{2}(\vc{p}_1-\vc{p}_2)$ is the 
relative two-nucleon momentum) \cite{DKA00,HF86} 
\begin{equation} 
\eta_2(\vc{P}_{\mathrm CM}) = \int  
   \eta_2(\vc{P}_{\mathrm CM}/2+\vc{q}_{\mathrm rel}, 
   \vc{P}_{\mathrm CM}/2-\vc{q}_{\mathrm rel}) {\d^3}q_{\mathrm rel}  ,   
\end{equation} 
\begin{equation}
\eta_2(\vc{q}_{\mathrm rel}) = \int 
\eta_2(\vc{P}_{\mathrm CM}/2+\vc{q}_{\mathrm rel},
\vc{P}_{\mathrm CM}/2-\vc{q}_{\mathrm rel}) {\d^3}P_{\mathrm CM}  . 
\label{Ecmrels} 
\end{equation} 
It is also related to the combined two-body center-of-mass and 
relative momentum distribution 
$\eta_2(P_{\mathrm CM},q_{\mathrm rel})$ 
\cite{OS95}  
\begin{equation}
\eta_2(P_{\mathrm CM},q_{\mathrm rel})=  
   \int \eta_2(\vc{P}_{\mathrm CM}/2+\vc{q}_{\mathrm rel}, 
   \vc{P}_{\mathrm CM}/2-\vc{q}_{\mathrm rel}) 
   {\d}\Omega_{P_{\mathrm CM}}{\d}\Omega_{q_{\mathrm rel}} . 
\label{Ecmrel}
\end{equation} 
Its relationship with the two-nucleon spectral function 
$S(\vc{p}_1,\vc{p}_2;E)$ is 
\begin{equation} 
\eta_2(\vc{p}_1,\vc{p}_2) = \int{{\d} E}  
S(\vc{p}_1,\vc{p}_2;E) . 
\label{Espectral}
\end{equation} 

In a totally uncorrelated system the 2bMD is factorized as 
\begin{equation} 
\eta_2(\vc{p}_1,\vc{p}_2) =  
   (1-\frac{1}{A})\eta (\vc{p}_1)\eta (\vc{p}_2). 
\label{Eus} 
\end{equation} 
The constant 
$1-1/A$ appears due to the normalization condition 
(\ref{Enor}). 
For the ground state of a non-interacting Fermi system 
(a system of $A$ identical, non-interacting fermions) 
where only statistical correlations are present, 
starting from the 2DM \cite{AHP88}, 
the 2bMD can be shown to have the form 
\begin{equation} 
\eta_2(\vc{p}_1,\vc{p}_2) = 
   \eta (\vc{p}_1)\eta (\vc{p}_2) 
   - \frac{1}{\nu}|\eta_1(\vc{p}_1,\vc{p}_2)|^2   ,  
\label{Efs} 
\end{equation} 
where $\eta_1(\vc{p}_1,\vc{p}_2)$ is the one-body 
density matrix 
in momentum space 
and $\nu$ is the level degeneracy. 
Note that in the special case of a non-interacting Fermi system 
with $A=\nu $ 
Eq.~(\ref{Efs}) 
becomes identical with Eq.~(\ref{Eus}). 

As an example we consider 
the ground state of 
an infinitely extended ideal Fermi gas 
(volume $\Omega$ and number of particles $A$ going to infinity 
while the particle density stays constant) 
with Fermi wave number $k_{\mathrm{F}}$ 
and degeneracy $\nu$. 
Equation~(\ref{Efs}) yields for the 2bMD per particle-pair 
$
\tilde{\eta}_2^{\mathrm{F}} (\vc{p}_1,\vc{p}_2) = 
\eta_2^{\mathrm{F}} (\vc{p}_1,\vc{p}_2) /[A(A-1)] 
$
\begin{equation} 
\tilde{\eta}_2^{\mathrm{F}} (\vc{p}_1,\vc{p}_2) = 
\left( 
\frac{4}{3}\pi k_{\mathrm F}^3 
\right)^{-2} 
\theta (k_{\mathrm{F}}-p_1)\theta(k_{\mathrm{F}}-p_2) 
[1 - \delta_{\vc{p}_1\vc{p}_2}/\nu ] . 
\label{En2fg} 
\end{equation} 
The momentum distribution per particle, 
$\tilde{\eta}^{\mathrm F} (p)=
\eta^{\mathrm F} (p)/A =\eta^{\mathrm F} (\vc{p})/A$,  
is given by 
\begin{equation} 
\tilde{\eta}^{\mathrm{F}}(p) = 
\left( 
\frac{4}{3}\pi k_{\mathrm F}^3 
\right)^{-1} 
\theta (k_{\mathrm{F}}-p) . 
\end{equation} 
Let us define the quantity $\xi$ 
as a measure of finite-size effects, of statistical and 
short-range correlations of dynamical origin 
\begin{equation} 
 \xi (\vc{p}_1,\vc{p}_2) \equiv 
   \eta_2(\vc{p}_1,\vc{p}_2) / 
   \eta (\vc{p}_1)\eta (\vc{p}_2) .  
\label{Exi} 
\end{equation} 
For an infinitely extended 
ideal Fermi gas 
$\xi$ is defined for 
$p_1,p_2\leq k_F$ and equals 
\begin{equation} 
\xi^{\mathrm{F}} (\vc{p}_1,\vc{p}_2) = 
	 \left\{  \begin{array}{ll} 
	  1, & \vc{p}_1\neq\vc{p}_2 \\ 
		  1-1/\nu , & \vc{p}_1 = \vc{p}_2 
                  \end{array} \right. \hspace{5mm}.  
\label{Exifg} 
\end{equation} 
From Eqs.~(\ref{Eus}) and (\ref{Exifg}) we realize that, 
in the case of the infinitely extended ideal Fermi system, 
if $\vc{p}_1\neq\vc{p}_2$,
$\xi$ equals 1. This holds even in the presence 
of dynamical SRC, if long-range order 
does not exist \cite{Fa81}. 
In the case of the infinitely extended ideal Fermi gas, 
$\xi =1-1/\nu$ if $\vc{p}_1=\vc{p}_2$. 
As for the finite, non-interacting Fermi system, 
we realize from Eq.~(\ref{Efs}) 
that, if $\vc{p}_1=\vc{p}_2$, $\xi =1-1/\nu$. 
Deviations of $\xi$ from this value 
show the effect of dynamical correlations. 
For $\vc{p}_1\neq\vc{p}_2$ deviations of $\xi$ from 
$1 - 1/A$ 
is a measure of statistical and (or) dynamical correlations 
in a system of finite size. 

\section{Method of Calculation for $\ell$-Closed Nuclei 
- Application to {$^{16}$O}} 
\label{Smoc} 
\subsection{Method of Calculation} 
\label{Smoc1} 
Let us now consider a system of 
$A$ identical non-interacting fermions
in its ground state. The fermions occupy the lowest 
single-particle
energy eigenstates 
$|n_j\rangle$ ($j=1,2,...,A/\nu$)  
and the 2bMD is given by Eq.~(\ref{Efs}).
In the case of the nucleus, 
$A$ is the mass number 
and $\nu$ is 
the degeneracy due to the nucleon spin and isospin.
We consider $Z=N$, $\ell$-closed nuclei. 
In order to obtain closed analytical expressions for 
$\eta_2 (\vc{p}_1,\vc{p}_2)$, 
we have 
assumed that the nucleons move in an isotropic 
harmonic oscillator potential 
and that 
the spin-orbit coupling is negligible, 
the center-of-mass 
and finite nucleon size corrections 
are small and 
the Coulomb interaction (relevant for protons) is small. 
Then Eq.~(\ref{Efs}) gives: 
\begin{equation} 
\eta_2 (\vc{p}_1,\vc{p}_2) = 
\frac{b^6}{\pi^{3}} 
\e^{-p_1^2b^2} 
\e^{-p_2^2b^2} 
\sum_{\mu_1 =0}^{2N_{\max}} (p_1 b)^{\mu_1} 
\sum_{\mu_2 =0}^{2N_{\max}} (p_2 b)^{\mu_2} 
{\mathcal{G}}_{\mu_1\mu_2}(\cos{\gamma}) ,  
\label{En2} 
\end{equation} 
where the coefficients ${\mathcal{G}}_{\mu_1\mu_2}$ 
are given by
\begin{equation} 
{\mathcal{G}}_{\mu_1\mu_2}(\cos{\gamma}) 
 = 4f_{\mu_1/2}f_{\mu_2/2}\! -\! 
   \sum_{\lambda_1=0}^{\mu_1}\! 
   \sum_{\lambda_2=0}^{\mu_2} 
   {\mathcal{K}}_{\lambda_1\lambda_2}(\cos{\gamma})  
   {\mathcal{K}}_{(\mu_1\!-\!\lambda_1)(\mu_2\!-\!\lambda_2)}(\cos{\gamma})  
\end{equation} 
and the coefficients $f_{\mu /2}$ 
and ${\mathcal{K}}_{\lambda_1\lambda_2}(\cos{\gamma})$ 
are given by 
\begin{equation} 
   f_{\mu /2}  
 = \sum_{n\ell,{\mbox{\scriptsize occ.}}} 
   f_{n\ell}^{\mu /2}, 
\label{Eff} 
\end{equation} 
\begin{equation} 
   {\mathcal{K}}_{\lambda_1\lambda_2}(\cos{\gamma})  
 = \sum_{n\ell,{\mbox{\scriptsize occ.}}}
   K_{n\ell}^{\lambda_1\lambda_2}P_{\ell}(\cos{\gamma}).  
\label{EKf} 
\end{equation} 
In the above equations 
$\gamma$ is the angle between the vectors 
$\vc{p}_1$ and $\vc{p}_2$, 
b is the harmonic oscillator parameter, 
$P_{\ell}(x)$ is a Legendre polynomial and  
$N_{\max}=(2n+\ell )_{\max}$ is the number of energy quanta 
of the highest occupied $n\ell$-level. 
The coefficients 
$f_{n\ell}^{\mu /2}$, $K_{n\ell}^{\lambda_1\lambda_2}$ 
are rational numbers that enter  
the corresponding expressions of 
$\eta(\vec{p})$ 
and 
$\eta_1(\vec{p}_1,\vec{p}_{1'})$ 
respectively. 
Analytical expressions 
for $f_{n\ell}^{\mu /2}$ 
and $K_{n\ell}^{\lambda_1\lambda_2}$ 
can be found 
in Ref.~\cite{PMK00}. 
They are different from zero 
only if the indices $\mu$ and $\lambda_1 +\lambda_2$ are even. 

The corresponding expression for the 
spherically symmetric nucleon momentum distribution 
$\eta (\vc{p})$ 
is \cite{KV90}  
\begin{equation} 
\eta (\vc{p}) = 
\eta (p) = 
\frac{b^3}{\pi^{3/2}} 
\e^{-p^2b^2} 
\sum_{\lambda =0}^{N_{\max}}  (p b)^{2\lambda} 
2f_{\lambda} . 
\label{Enp} 
\end{equation} 
It is verified that 
the 2bMD $\eta_2 (\vc{p}_1,\vc{p}_2)$ 
given by Eq.~(\ref{En2}) 
satisfies the property 
of Eq.~(\ref{Eseq}).

\subsection{Application to {$^{16}$O}} 
\label{Smoc2} 
Using the expressions 
(\ref{En2}) and (\ref{Enp}), 
we obtain results 
for the quantities 
$\eta_2(\vc{p}_1,\vc{p}_2)$ and $\xi(\vc{p}_1,\vc{p}_2)$, 
in the case of the nucleus $^{16}$O. 
The harmonic oscillator parameter 
$b=1.7825$fm 
has been determined so as to reproduce 
the experimental value of the 
charge root mean square radius of $^{16}$O, 
$\langle r\rangle^{1/2}=2.737$fm \cite{VJ87}. 
In Fig.~\ref{f1} the 2bMD of \nuc{16}{O} in the harmonic oscillator model 
(continuous lines) 
is plotted 
for $\vc{p}_1$ parallel to $\vc{p}_2$ 
and for $p_1=$~0,~1 and 1.5 fm$^{-1}$, 
as a function of the variable p$_2$ 
($\vc{p}_2={\mathrm p}_2\hat{p}_1$). 
In the same graphs the 2bMD 
of an infinite ideal Fermi gas 
is shown (dashed lines), 
calculated via 
Eq.~(\ref{En2fg}) with $\nu =4$ and 
with Fermi wave number $k_{\mathrm{F}}$ 
equal to the one corresponding to \nuc{16}{O}, 
namely $k_{\mathrm F}=1.1$fm$^{-1}$. 
This latter value has been calculated using local density approximation 
and coincides with the one used in Ref.~\cite{PWP92}. 
In order to make the comparison meaningful, 
the 2bMD of the ideal Fermi gas 
was normalized to the same number of particle pairs as 
those found in \nuc{16}{O}, 
namely to $A(A-1)=240$. 

It is evident from Eq.~(\ref{En2fg}) 
that the 2bMD of the ideal Fermi gas 
exhibits discontinuities 
at $p_{1,2}=k_{\mathrm{F}}$ and at $\vc{p}_1=\vc{p}_2$. 
This is observed in the first two panels of Fig.~\ref{f1} 
($p_1=$~0,~1~fm$^{-1}$) 
at p$_2=p_1$ and at p$_2=\pm k_{\mathrm{F}}$. 
In the case of the finite system \nuc{16}{O} 
in the harmonic oscillator model 
a rather similar behaviour is observed, 
but the 2bMD changes smoothly 
with p$_2$. 
For $p_1=1.5$fm$^{-1}$ 
(which is higher than $k_{\mathrm{F}}=1.1$fm$^{-1}$) 
the 2bMD of the ideal Fermi gas has vanished and that of \nuc{16}{O} 
has dropped considerably. 

In Fig.~\ref{f2}  
the quantity $\xi (\vc{p}_1,\vc{p}_2)$ 
is plotted in the case of \nuc{16}{O} 
as a function of $\cos{\gamma}$ 
and for 
$p_1=p_2=$~0,~1 and 4 fm$^{-1}$. 
The values of $\xi$ 
lie between 0.75($=1-1/{\nu}$) and 1. 
$\xi$ is equal to 0.75 when $\vc{p}_1=\vc{p}_2$ 
(as shown for example in the first panel, 
and in the other two panels when $\cos{\gamma}=1$). 
We realize that for 
high values of $p_1$ and $p_2$, 
$\xi$ tends to 1 and 0.75 
for $\cos{\gamma}=0$ and $\cos{\gamma}=-1$ respectively. 
The deviation of the values of $\xi$ 
from 1 or 0.75 is a finite-size effect 
in the presence of statistical correlations. 
In the case when $p_1\neq p_2$ (not shown) 
the quantity $\xi$ as a function of $\cos\gamma$ 
remains higher than 0.75 and below or equal to 1, 
following a similar behaviour as in Fig.~\ref{f2}, 
namely $\xi (-1)>\xi (1)$ and $\xi (x_0)=1$ 
for some value $\cos\gamma =x_0<0$. 

\section{The Effect of Short-Range Correlations} 
\label{Sdc} 
Up to now 
we have ignored the effect of dynamical, 
short-range correlations 
on the 2bMD. 
It is expected that 
such a calculation 
is reliable only in restricted kinematical domains, 
namely for $p_1$, $p_2\leq k_{\mathrm{F}}$. 
In this Section we make an attempt 
to investigate the effect 
of the SRC 
by including Jastrow-type correlations 
in our study. 

\subsection{Method of Calculation} 
\label{Smocc} 
The approach adopted here is based 
on the Jastrow formalism with a 
state-independent central correlation function, 
which introduces the 
short range correlations. 
Employing the low-order approximation (LOA) 
of Ref.~\cite{DSB82} 
for the 2DM 
and performing the spin-isospin summation ($\nu =4$) 
in Eq.~(14) of Ref.~\cite{DSB82}, 
we obtain for the 
correlated 2DM 
$\rho_2^{\mathrm{corr}} 
(\vc{r}_1,\vc{r}_2;\vc{r}_{1'},\vc{r}_{2'})$ 
the following expression:
\begin{eqnarray} 
\!\!\!\!&\rho_2^{\mathrm{corr}}&
\!\!(\vc{r}_1,\vc{r}_2;\vc{r}_{1'},\vc{r}_{2'}) = 
[1+g(r_{12},r_{1'2'})] 
\rho_2(\vc{r}_1,\vc{r}_2;\vc{r}_{1'},\vc{r}_{2'})  
    \nonumber \\ 
& & 
  +\int [g(r_{13},r_{1'3})+g(r_{23},r_{2'3})] 
  [\rho_1(\vc{r}_1,\vc{r}_{1'})\rho_{2h}(\vc{r}_2,\vc{r}_3,\vc{r}_{2'})  
    \nonumber \\ 
& & 
  -\nu^{-1}
  \rho_1(\vc{r}_1,\vc{r}_{2'})\rho_{2h}(\vc{r}_2,\vc{r}_3,\vc{r}_{1'})  
  -\nu^{-1}
  \rho_1(\vc{r}_1,\vc{r}_3)\rho_2(\vc{r}_2,\vc{r}_3;\vc{r}_{2'},\vc{r}_{1'})  
  ] {\d^3}r_3 
    \nonumber \\ 
& & 
  -\nu^{-1} 
  \int\int 
  g(r_{34},r_{34}) 
  \{ \rho_{2h}(\vc{r}_2,\vc{r}_4,\vc{r}_3) 
     \rho_2(\vc{r}_1,\vc{r}_3;\vc{r}_{1'},\vc{r}_{2'}) 
    \nonumber \\ 
& & 
  + \rho_1(\vc{r}_1,\vc{r}_3) 
    [ \rho_1(\vc{r}_2,\vc{r}_{2'}) \rho_{2h}(\vc{r}_3,\vc{r}_4,\vc{r}_{1'})
  -\nu^{-1}
  \rho_1(\vc{r}_2,\vc{r}_{1'})\rho_{2h}(\vc{r}_3,\vc{r}_4,\vc{r}_{2'})  
    \nonumber \\ 
& & 
  -\nu^{-1}
  \rho_1(\vc{r}_2,\vc{r}_4)\rho_2(\vc{r}_3,\vc{r}_4;\vc{r}_{1'},\vc{r}_{2'})  
  ]\}  {\d^3}r_3 
    {\d^3}r_4 
 ,\label{Er2c} 
\end{eqnarray} 
where 
$r_{ij}=|\vc{r}_i-\vc{r}_j|$,   
$g(r,r')\equiv f(r)f(r')-1$ 
and $f(r)$ is the correlation function, 
which has to obey the conditions 
$f(0)<1$ and for $r\rightarrow\infty$, $f(r)\rightarrow 1$.

In Eq.~(\ref{Er2c}) 
the 
one-body density matrix 
$\rho_1(\vc{r},\vc{r'})$, 
the 2DM 
$\rho_2(\vc{r}_1,\vc{r}_2;\vc{r}_{1'},\vc{r}_{2'})$ 
and the half-diagonal 2DM 
$\rho_{2h}(\vc{r}_1,\vc{r}_2,\vc{r}_{1'})$ 
are calculated in the harmonic-oscillator model. 
The correlated 2bMD, $\eta_2^{\mathrm{corr}} (\vc{p}_1,\vc{p}_2)$, 
is then calculated by Fourier transforming 
$\rho_2^{\mathrm{corr}}(\vc{r}_1,\vc{r}_2;\vc{r}_{1'},\vc{r}_{2'})$ 
according to Eq.~(\ref{Edft}). 
The correlated momentum distribution 
$\eta^{\mathrm{corr}}(\vc{p})$ 
is calculated likewise 
in the LOA 
by performing the spin-isospin summation 
in expression (13) of Ref.\,\cite{DSB82} 
for the one-body density matrix 
and by Fourier-transforming with respect to 
$\vc{r}_1-\vc{r}_{1'}$. 

The LOA preserves 
the normalization of the density matrices. 
As a consequence, 
the one- and two-body momentum distributions 
calculated in the LOA obey the normalization conditions 
(\ref{Enor1}) and (\ref{Enor}), 
as well as the sequential relation (\ref{Eseq}). 
However, in the LOA 
some probability distributions 
(i.e. positive-definite quantities) 
may obtain negative values within regions of their domain. 
This problem is probably related to the 
$A-$representability of the LOA two-body 
density matrices.
For example, it was encountered in 
calculations of the relative distribution function 
$\rho_2(r_{12})$ \cite{DSB82} 
and the combined two-body center-of-mass and 
relative momentum distribution 
$\eta_2(P_{\mathrm CM},q_{\mathrm rel})$ 
\cite{OS95}, 
which implies that it may show up 
in the 2bMD too, since the latter 
is related to 
$\eta_2(P_{\mathrm CM},q_{\mathrm rel})$ 
via Eq.~(\ref{Ecmrel}). 
In the case of the $\rho_2(r_{12})$ 
the problem was subsequently solved 
by an appropriate modification of the correlation function 
\cite{DKA00}. 
The above remarks are highly relevant 
to the choice of the correlation function 
used in this work, 
as we will discuss in due course. 


\subsection{Application to \nuc{4}{He}} 
\label{Sahe} 
The 2bMD 
and the momentum distribution 
of \nuc{4}{He} 
in the harmonic oscillator model, 
which we will hereafter denote by 
$\eta_2^0(\vc{p}_1,\vc{p}_2;b) $    
and $\eta^0 (p;b)$ respectively,   
are given by 
(see Eqs.~(\ref{En2}),~(\ref{Enp}))  
\begin{equation} 
\eta_2^0(\vc{p}_1,\vc{p}_2;b) =  
\frac{12b^6}{\pi^3} 
\e^{ -b^2[p_1^2+p_2^2]} , \label{En2H}   
\end{equation} 
\begin{equation} 
\eta^0 (p;b) = \frac{4b^3}{\pi^{3/2}} \e^{-p^2b^2} . 
\end{equation} 
yielding $\xi (\vc{p}_1,\vc{p}_2)=0.75$. 
The case falls into the category of 
non-interacting Fermi systems with $A=\nu $,
hence Eq.~(\ref{Eus}) holds 
(see remark under Eq.~(\ref{Efs})). 
The harmonic oscillator parameter 
$b=1.382$fm 
reproduces the experimental value of the 
charge root mean square radius of $^{4}$He,  
$\langle r^2\rangle^{1/2}=1.67$fm 
\cite{VJ87}  
in this model (corrections due to center-of-mass motion 
and finite nucleon size are taken into account). 
We include SRC in the LOA using 
a two-gaussian (2G) correlation function, 
$f(r)=1-c_1\exp (-r^2/\beta_1^2) 
+c_2\exp (-r^2/\beta_2^2)$. 
The expression of the 2bMD is 
\begin{eqnarray} 
& &\!\!\!\!\eta_2^{\mathrm{corr}}(\vc{p}_1,\vc{p}_2)  
  = \eta_2^0 (\vc{p}_1,\vc{p}_2;b)  
+ \delta{\eta}_2(\vc{p}_1,\vc{p}_2;b,c_1,y_1) 
+ \delta{\eta}_2(\vc{p}_1,\vc{p}_2;b,-c_2,y_2) 
  \nonumber \\ 
&-&  
 24c_1c_2\frac{b^6}{\pi^3} 
\{  
    \frac{1}{ [(1+4y_1)(1+4y_2)]^{3/2}}
    \e^{ -b^2[\frac{1}{2}(\frac{1+2y_1}{1+4y_1}
	      +\frac{1+2y_2}{1+4y_2})(p_1^2+p_2^2) + 
             (\frac{2y_1}{1+4y_1} 
	      +\frac{2y_2}{1+4y_2})\vc{p}_1\cdot\vc{p}_2] }   
  \nonumber \\ 
&+&  
   \frac{2}{(1+3y_1+3y_2+8y_1y_2)^{3/2}} 
    [\e^{-\frac{1+2y_1+2y_2}{1+3y_1+3y_2+8y_1y_2}p_1^2b^2} 
     \e^{-p_2^2b^2} + (p_1 \leftrightarrow p_2) ]
  \nonumber \\ 
&-&  
   \frac{5}{(1+2y_1+2y_2)^{3/2}} \e^{-b^2(p_1^2+p_2^2)}  
\} 
\label{Eec2} 
\end{eqnarray}
where $y_i\equiv b^2/\beta_i^2$ and  
\begin{eqnarray} 
\delta{\eta}_2  
	       (\vc{p}_1,\vc{p}_2;b,c,y) &\equiv& 
\frac{12b^6}{\pi^3}\{ 
    [\frac{10c}{(1+2y)^{3/2}} - \frac{5c^2}{(1+4y)^{3/2}}]  
    \e^{ -b^2(p_1^2+p_2^2) } 
    \nonumber \\  
&-& 
    \frac{2c}{(1+4y)^{3/2}}
    \e^{ -b^2[\frac{1+3y}{1+4y}(p_1^2+p_2^2) + 
    \frac{2y}{1+4y}\vc{p}_1\cdot\vc{p}_2] }   
    \nonumber \\ 
&+&  
    \frac{c^2}{ (1+4y)^{3/2}} 
    \e^{ -b^2[\frac{1+2y}{1+4y}(p_1^2+p_2^2) + 
    \frac{4y}{1+4y}\vc{p}_1\cdot\vc{p}_2] }  
    \nonumber \\ 
&+&  
    \frac{2c^2}{[(1+2y)(1+4y)]^{3/2}} 
    [\e^{-\frac{1}{1+2y}p_1^2b^2}\e^{-p_2^2b^2} 
    + (p_1\leftrightarrow p_2)] 
    \nonumber \\ 
&-& \frac{4c}{(1+3y)^{3/2}} 
    [\e^{-\frac{1+2y}{1+3y}p_1^2b^2}\e^{-p_2^2b^2} 
    + (p_1\leftrightarrow p_2)] \}
   . \label{En2Hc} 
\end{eqnarray} 
The corresponding expression for the 
momentum distribution 
$\eta^{\mathrm corr}(p)$ 
of \nuc{4}{He} is 
\begin{eqnarray} 
& & \eta^{\mathrm corr}(p) = 
   \eta^0(p;b) +\delta\eta(p;b,c_1,y_1) + \delta\eta(p;b,-c_2,y_2) 
 \nonumber 
 \\ 
 & & 
   -6c_1c_2\frac{b^3}{\pi^{3/2}} 
    \left[  \frac{
     \e^{-p^2b^2\frac{1+2y_1+2y_2}{1+3y_1+3y_2+8y_1y_2}} }{ 
    (1+3y_1+3y_2+8y_1y_2)^{3/2}  } 
   -\frac{\e^{-p^2b^2}}{ 
   (1+2y_1+2y_2)^{3/2}}
   \right], 
\label{Eec1} 
\end{eqnarray} 
where 
\begin{equation} 
\delta\eta (p;b,c,y)\equiv \frac{12b^3}{\pi^{3/2}} 
\left[\frac{ c^2{\e}^{-p^2b^2\frac{1}{1+2y}}}{ [(1+2y)(1+4y)^{3/2}} 
  -\frac{ 2c {\e}^{-p^2b^2\frac{1+2y}{1+3y}} }{(1+3y)^{3/2}}  
  \right]. 
\label{Ede1} 
\end{equation} 
If we put $c_2=0$ in the above equations we end up with the 
corresponding expressions for the single gaussian (1G) 
correlation function, $f(r)=1-c\exp{(-r^2/\beta^2)}$ 
which has been used in evaluating several quantities 
of \nuc{4}{He} \cite{BS80,DSB82,SAD93b,MM99,DKA00,OS95}.
We have introduced the 2G correlation function, 
that can overshoot unity at intermediate distances. 
This was found necessary in order to eliminate the negative values 
of $\eta_2(\vc{p}_1,\vc{p}_2)$ 
that result at some regions of intermediate values of momenta 
whenever a 1G correlation function is used. 

The parameterization of the new, 2G correlation function 
by means of a rigorous procedure such as 
minimization and fitting to some set of experimental data 
(e.g. charge form factor data) 
lies beyond the scope of this work, 
therefore it was obtained 
somehow heuristically. 
First, it has been chosen to maintain 
the small-$r$ behaviour of the 
1G correlation function 
of Ref.~\cite{DKA00},  
by setting $c_1-c_2=0.8$ and $\beta_1=0.8$fm. 
Subsequently, 
the extra parameters 
have been determined requiring that 
the healing distance 
is equal to or somewhat lower than 1fm and 
the relative pair density distribution 
$\rho_2(r_{12})$ is approximately zero 
at $r_{12}=0$. 
The value of the harmonic oscillator parameter $b$ 
was again chosen so as to reproduce the experimental 
charge root mean square radius of \nuc{4}{He}. 
We ended up with $c_2=1.7$ (and therefore $c_1=0.9$), 
$\beta_2=1.1$fm and $b=1.4$fm. 
In the left panel of Fig.~\ref{f3} the resulting 
2G correlation function, that overshoots unity, 
is plotted along with the 1G one of Ref.~\cite{DKA00}.
The corresponding values of a wound parameter 
$\kappa =\int (1-f(r_{12}))^2 \rho_2^0(r_{12})\d^3r_{12}$ 
(where $\rho_2^0(r_{12})$ is the 
relative pair density distribution function 
calculated in the 
harmonic oscillator model and normalized to unity) 
for the 2G and 1G correlations functions 
are equal to 0.008 and 0.018 respectively. 
In the center panel of the same figure 
the momentum distribution per particle 
$\eta (p)/A$ 
is plotted 
for both correlation functions, 2G and 1G, along with 
the result of a variational calculation \cite{PWP92,Wi91}.
In the right panel of Fig.~\ref{f3} the proton-proton relative pair 
density distribution (normalized to unity) $\rho_{pp}(r_{12})$  
is shown for both correlation functions 2G, 1G. 
The bars cover the results of several variational calculations 
using a variety of two- and three-nucleon interaction models 
\cite{Ca88b}. 
The values of $\eta (p)$ 
and $\rho_{pp}(r_{12})$ 
calculated with the two-gaussian correlation function 
compare quite well with the ones of other methods. 

Some numerical results for $\eta_2(\vc{p}_1,\vc{p}_2)$ 
are presented in Figs.~\ref{f4} and \ref{f5}. 
In Fig.~\ref{f4}, the 2bMD of \nuc{4}{He} is plotted 
in logarithmic scale, 
   for $\vc{p}_2$ parallel to $\vc{p}_1$, $\vc{p}_2=$p$_2\hat{p}_1$,  
   as a function of p$_2$, 
   for $p_1=$~0,~1,~2,~3~fm$^{-1}$. 
   The correlated 2bMD, as given by Eq.~(\ref{Eec2}), 
   is plotted with continuous lines 
   and the harmonic-oscillator result, 
   Eq.~({\ref{En2H}), 
   with dashed lines. 
   It seems that, within the present approximation, 
   the correlated 2bMD shows deviations 
   from the harmonic-oscillator picture 
   mainly at high values of $p_1$ and/or $p_2$, 
developing high-momentum tails. 
In most cases, 
the deviations are larger when $\vc{p}_1$, $\vc{p}_2$ are antiparallel 
rather than parallel. 

In Fig.~\ref{f5} 
   the quantity $\xi (\vc{p}_1,\vc{p}_2)$ 
   for $^4$He 
   is plotted 
   in logarithmic scale, 
   as a function of $\cos{\gamma}$ 
   for $p_1=p_2=$~2 and 4~fm$^{-1}$. 
   The continuous line includes SRC and the dashed line corresponds 
   to the harmonic oscillator model and it equals 0.75. 
   Deviations from this value shows the effect of SRC. 
The correlated $\xi$ in the present approximation 
grows larger than unity 
at backward angles, 
and more so for large values of $p_1$, $p_2$, as indicated by the 
$p_1=p_2=4$fm$^{-1}$ graph. 
In this latter graph, 
deviations from 0.75 are large even at forward angles. 
We should note that for 
low values of $p_1$ and $p_2$ 
(not shown), 
the correlated $\xi$ takes values between about 0.7 and 1.

We have also compared our results for the 2bMD 
with those 
obtained by using the single-gaussian 
correlation function of Ref.~\cite{DKA00}. 
Although the overall behaviour of the 2bMD 
is not dramatically affected by the 
choice of $f(r)$, 
the two sets of results differ largely 
at certain kinematical regions, 
in particular at intermediate values of momenta 
where the single-gaussian correlation function yields
negative results. 
(If the 2DM is approximated by only the first terms 
of the expression (\ref{Er2c}), 
i.e. $\rho_2^{\mathrm corr}=f(r_{12})f(r_{1'2'})\rho_2$, 
these regions of negative values show up 
as exact zeros at certain points \cite{PMK00a}.)  
As anticipated in Sec.~\ref{Smocc},  
a similar, unphysical situation has been encountered in 
the study of other two-body quantities, 
when the LOA is used \cite{DSB82,OS95}. 
It has been argued \cite{OS95} 
that such regions indicate 
where the SRC start to show up dramatically, 
and are most sensitive to the 
approximation employed to account for the SRC. 
The results  in these regions are also sensitive to 
the particular correlation function used. 
Let us note here that even in the case of the 
2G correlation function 
a somewhat lower value of the 
oscillator parameter (eg. $b=1.35$fm) 
retrieves the negative values of the 2bMD, 
indicating that the  
uncorrelated single-particle wavefunctions 
also play some role. 

\section{Summary and Conclusions} 
\label{Ssap} 
In summary, 
a compact analytical expression has been derived 
for the two-body momentum distribution, 
$\eta_2 (\vc{p}_1,\vc{p}_2)$, 
of $Z=N$, $\ell$-closed nuclei 
within the context of the independent-particle shell model, 
using harmonic-oscillator wavefunctions. 
An application to the \nuc{16}{O} nucleus 
revealed interesting features arising from the 
Fermi statistics and the finite nuclear size. 
Next, the effect of short-range correlations 
was investigated by including Jastrow-type correlations in the 
calculation of 
$\eta_2 (\vc{p}_1,\vc{p}_2)$ 
of the \nuc{4}{He} nucleus, 
using the low-order approximation  
of Ref.~\cite{DSB82} 
and a two-gaussian correlation function. 
Significant deviations from the independent-particle picture 
were found for large values of $p_1$ and $p_2$ and 
for angles 
between the vectors 
$\vc{p}_1$ and $\vc{p}_2$ 
close to $\gamma =$180$^{\circ}$ and 0$^{\circ}$. 
The quality of our approximation was also explored. 
It was found that, at certain regions of momenta, 
the results are sensitive to 
the correlation function used. 

Using the same approximation, 
one could evaluate the 2bMD and investigate 
the effects of short-range correlations 
in other interesting nuclei, 
for example in \nuc{16}{O} and \nuc{40}{Ca}. 
Furthermore, a general expression 
within the harmonic oscillator model 
applicable to every $\ell -$closed nucleus 
could be derived for  
the two-body momentum distribution 
$\eta_2 (\vc{p}_1,\vc{p}_2)$, 
the two-body center-of-mass momentum distribution 
$\eta_2(\vc{P}_{\mathrm CM})$ 
and the 
two-body relative momentum distribution 
$\eta_2(\vc{q}_{\mathrm rel})$, 
as well as for the
two-hole spectral function $S(\vc{p}_1,\vc{p}_2;E)$.  
The above calculations are relevant to the study 
of the two-nucleon knock-out reactions. \\[5mm] 

\noindent 
{\bf Acknowledgements}\\[0.5mm] 

\noindent 
Partial financial support from the University of Athens 
under Grant 70/4/3309 is acknowledged. 
P.P. acknowledges a scholarship from the 
Greek State Scholarship Foundation (IKY). 



\begin{thebibliography}{10}

\bibitem{AHP93}
{A.N.\,Antonov}, {P.E.\,Hodgson}, and {I.Zh.\,Petkov}.
\newblock {\em Nucleon Correlations in Nuclei}.
\newblock Springer--Verlag, 1993;
%
{L.S.\,Cardman}.
\newblock {\em Nucl. Phys. {\bf A654}}, (1999) 73c.

\bibitem{La93}
{L.\,Lapikas}.
\newblock {\em Nucl. Phys. {\bf A553}}, (1993) 297c.

\bibitem{PS97}
{V.R.\,Pandharipande}, {I.\,Sick}, and {P.K.A.\,de Witt Huberts}.
\newblock {\em Rev. Mod. Phys. {\bf 69}}, (1997) 981.

\bibitem{AHP88}
{A.N.\,Antonov}, {P.E.\,Hodgson}, and {I.Zh.\,Petkov}.
\newblock {\em Nucleon Momentum and Density Distributions in Nuclei}.
\newblock Claredon Press, Oxford, 1988.

\bibitem{MP00}
{H.\,M\"uther} and {A.\,Polls}.
\newblock {\em Prog. Part. Nucl. Phys. {\bf 45}}, (2000) 243.

\bibitem{BF90}
{\em {{\rm Proceedings of the Workshop on}} Two-Nucleon Emission Reactions,
  {{\rm Elba International Physics Center, September 1989}}}.
\newblock Eds. O.\,Benhar and A.\,Fabrocini. ETS Editrice, Pisa, 1990.

\bibitem{Ja95}
{E.\,Jans}.
\newblock {\em {{\rm Proceedings of the Second Workshop on}}
  Electromagnetically Induced Two-Nucleon Emission}, Gent, May 17-20 1995,
  p.47.

\bibitem{Ro00}
{G.\,Rosner}.
\newblock {\em Prog. Part. Nucl. Phys. {\bf 44}}, (2000) 99 and references
  therein.

\bibitem{St00}
{R.\,Starink} et~al.
\newblock {\em Phys. Lett. {\bf B474}}, (2000) 33.

\bibitem{Ne98}
\newblock {MAMI proposals, A1 collaboration}, 
{R.\,Neuhausen} et~al. 
A1/1-97, A1/4-98, A1/5-98.   

\bibitem{Ba91}
\newblock {JLab proposals}, 
{E.\,Baghei} et~al. (spokespersons) 
PR-91-009, 
{W.\,Bertozzi} et~al. (spokespersons) 
E-01-015.  

\bibitem{Wa00}
{D.P.\,Watts} et~al.
\newblock {\em Phys. Rev. {\bf C62}}, (2000) 014616;
%
{C.J.Y.\,Powrie} et~al.
\newblock {\em Phys. Rev. {\bf C64}}, (2001) 034602.

\bibitem{Ac99}
{J.\,Aclander} et~al.
\newblock {\em Phys. Lett. {\bf B453}}, (1999) 211;
%
{A.\,Tang} et~al.
\newblock {\em nucl-ex/0206003}, submitted to Phys. Rev. Lett.

\bibitem{BGP96}
{S.\,Boffi}, {C.\,Giusti}, {F.D.\,Pacati}, and {M.\,Radici}.
\newblock {\em Electromagnetic Response of Atomic Nuclei}.
\newblock Claredon Press, Oxford, 1996.

\bibitem{Ge96}
{W.J.W.\,Geurts} et~al.
\newblock {\em Phys. Rev. {\bf C54}}, (1996) 1144.

\bibitem{Gi98}
{C.\,Giusti} et~al.
\newblock {\em Phys. Rev. {\bf C57}}, (1998) 1691.

\bibitem{Ry97}
{J.\,Ryckebusch} et~al.
\newblock {\em Nucl. Phys. {\bf A624}}, (1997) 581.

\bibitem{PMC01}
{M.\,Petraki}, {E.\,Mavrommatis}, and {J.W.\,Clark}.
\newblock {\em Phys. Rev. {\bf C64}}, (2001) 024301.

\bibitem{MPC95}
{E.\,Mavrommatis}, {M.\,Petraki}, and {J.W.\,Clark}.
\newblock {\em Phys. Rev. {\bf C51}}, (1995) 1849.

\bibitem{PMK00}
{P.\,Papakonstantinou}, {E.\,Mavrommatis}, and {T.S.\,Kosmas}.
\newblock {\em Nucl. Phys. {\bf A673}}, (2000) 171.

\bibitem{BF00}
{O.\,Benhar} and {A.\,Fabrocini}.
\newblock {\em Phys. Rev. {\bf C62}}, (2000) 034304.

\bibitem{KV90}
{T.S.\,Kosmas} and {J.D.\,Vergados}.
\newblock {\em Nucl. Phys. {\bf A510}}, (1990) 641.

\bibitem{KV92}
{T.S.\,Kosmas} and {J.D.\,Vergados}.
\newblock {\em Nucl. Phys. {\bf A536}}, (1992) 72.

\bibitem{Pa95e}
{P.\,Papakonstantinou}.
\newblock {Diploma thesis}, {University of Athens}, 1995.

\bibitem{GGR71}
{M.\,Gaudin}, {J.\,Gillespie}, and {G.\,Ripka}.
\newblock {\em Nucl. Phys. {\bf A176}}, (1971) 237.

\bibitem{BS80}
{O.\,Bohigas} and {S.\,Stringari}.
\newblock {\em Phys. Lett. {\bf 95B}}, (1980) 9.

\bibitem{DSB82}
{M.\,Dal R\`i}, {S.\,Stringari}, and {O.\,Bohigas}.
\newblock {\em Nucl. Phys. {\bf A376}}, (1982) 81.

\bibitem{SAD93b}
{M.V.\,Stoitsov}, {A.N.\,Antonov}, and {S.S.\,Dimitrova}.
\newblock {\em Phys. Rev. {\bf C48}}, (1993) 74.

\bibitem{MM99}
{S.E.\,Massen} and {Ch.C.\,Moustakidis}.
\newblock {\em Phys. Rev. {\bf C60}}, (1999) 024005.

\bibitem{MM00}
{Ch.C.\,Moustakidis} and {S.E.\,Massen}.
\newblock {\em Phys. Rev. {\bf C62}}, (2000) 034318.

\bibitem{DKA00}
{S.S.\,Dimitrova}, {D.N.\,Kadrev}, {A.N.\,Antonov}, and {M.V.\,Stoitsov}.
\newblock {\em Eur.Phys.J. {\bf A7}}, (2000) 335.

\bibitem{OS95}
{G.\,Orlandini} and {L.\,Sarra}.
\newblock {\em {{\rm Proceedings of the Second Workshop on}}
  Electromagnetically Induced Two-Nucleon Emission}, Gent, May 17-20 1995, p.1.

\bibitem{HF86}
{Y.\,Haneishi} and {T.\,Fujita}.
\newblock {\em Phys. Rev. {\bf C33}}, (1986) 260.

\bibitem{Fa81}
{S.\,Fantoni}.
\newblock {\em Nucl. Phys. {\bf A363}}, (1981) 381.

\bibitem{VJ87}
{H. de Vries}, {C.W. de Jager}, and {C. de Vries}.
\newblock {\em At. Dat. Nucl. Dat. Tables {\bf 36}}, (1987) 495.

\bibitem{PWP92}
{S.C.\,Pieper}, {R.B.\,Wiringa}, and {V.R.\,Pandharipande}.
\newblock {\em Phys. Rev. {\bf C46}}, (1992) 1741.

\bibitem{Wi91}
{R.B.\,Wiringa}.
\newblock {\em Phys. Rev. {\bf C43}}, (1991) 1585.

\bibitem{Ca88b}
{J.\,Carlson} et~al.
\newblock {\em Phys. Rev. {\bf C38}}, (1988) 1879.

\bibitem{PMK00a}
{P.\,Papakonstantinou}, {E.\,Mavrommatis}, and {T.S.\,Kosmas}.
\newblock {\em Proceedings of the 11th Symposium of the Hellenic Nuclear
  Physics Society}, Thessaloniki, Greece, 2000, to be published.

\end{thebibliography}

\clearpage 

\begin{figure} 
\centerline{\epsfig{figure=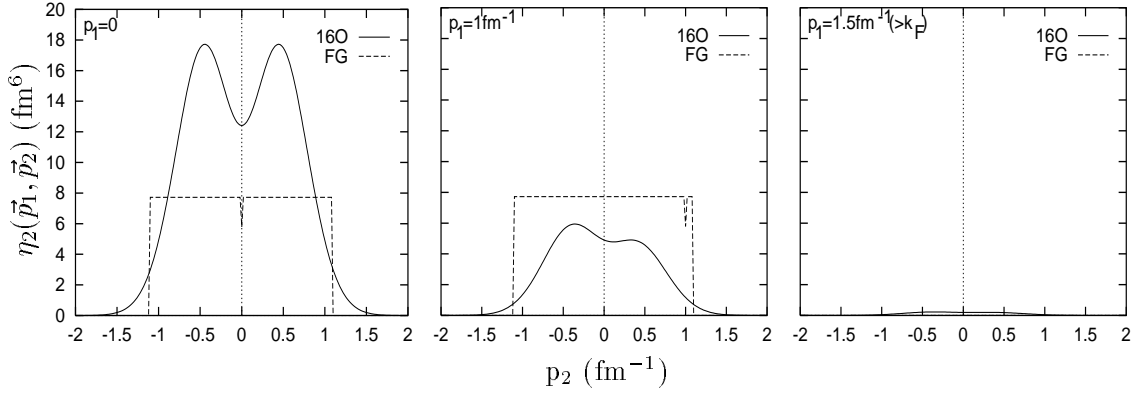}} 
\caption{ 
   Two-body momentum distribution 
   for $\vc{p}_1$ parallel to $\vc{p}_2$, 
   as a function of p$_2$ ($\vc{p}_2=$p$_2\hat{p}_1$) 
   for $p_1=$~0,~1,~1.5~fm$^{-1}$, 
   in the case of $^{16}$O 
   in the harmonic oscillator model 
   (continuous line) 
   and of an infinite ideal Fermi gas with 
   $k_{\mathrm{F}}=$1.1fm$^{-1}$, normalized to the same 
   number of particle pairs 
   as in $^{16}$O 
   (dashed line).}  
\label{f1} 
\end{figure} 
\begin{figure} 
\centerline{\epsfig{figure=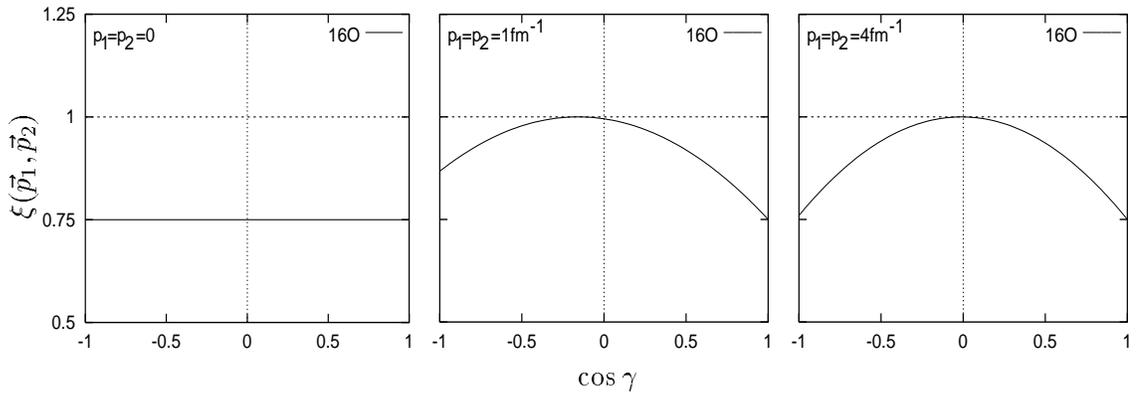}} 
\caption{ 
   The quantity $\xi (\vc{p}_1,\vc{p}_2)$ 
   (see Eq.~(\ref{Exi})) 
   as a function of $\cos{\gamma}$ 
   ($\gamma$ is the angle between the vectors $\vc{p}_1$ and $\vc{p}_2$) 
   for $p_1=p_2=$~0,~1,~4~fm$^{-1}$, 
   in the case of $^{16}$O in the harmonic oscillator model 
   (continuous line). 
} 
\label{f2} 
\end{figure} 
\begin{figure} 
\centerline{\epsfig{figure=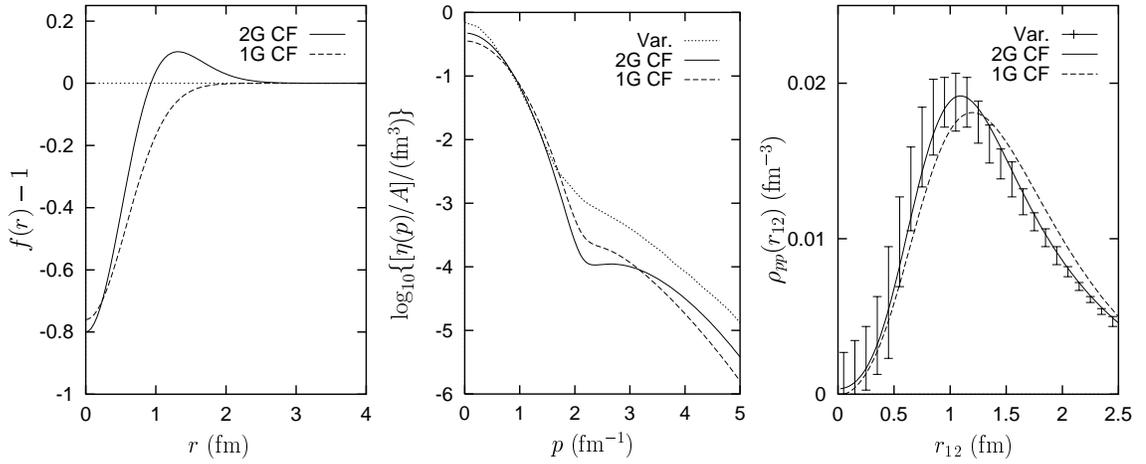,width=15cm}} 
\caption{ 
   Left: The single-gaussian correlation function 
   with $c=0.76$, $\beta =0.813$fm, from Ref.~\cite{DKA00} 
   (1G CF) and the two-gaussian correlation function 
   employed in this work with the parameterization 
   described in the text (2G CF). 
   Center: The momentum distribution per particle 
   of \nuc{4}{He} calculated in LOA with the 1G and 2G 
   correlation functions is compared with the variational 
   calculation of Ref.~\cite{PWP92} (``Var."). 
   Right: The proton-proton relative pair density 
   distribution (normalized to 1) of \nuc{4}{He} 
   calculated in LOA with the 1G and 2G correlation functions 
   is compared with the domain of results obtained via 
   a set of variational calculations \cite{Ca88b} (``Var."). 
} 
\label{f3} 
\end{figure} 
\begin{figure} 
\centerline{\epsfig{figure=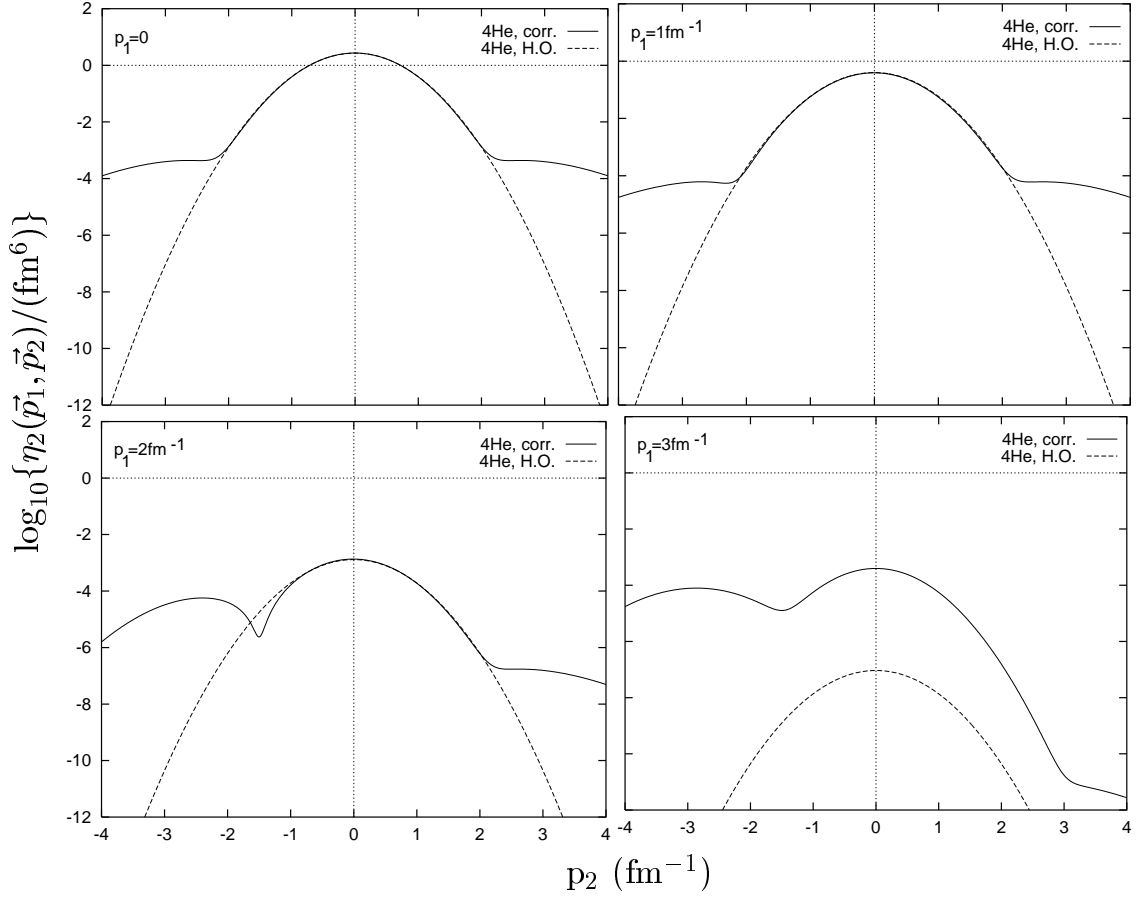,width=15cm}} 
\caption{ 
   Two-body momentum distribution 
   (in logarithmic scale) 
   for $\vc{p}_1$ parallel to $\vc{p}_2$, 
   as a function of p$_2$ ($\vc{p}_2=$p$_2\hat{p}_1$) 
   for $p_1=$~0,~1,~2,~3~fm$^{-1}$, 
   in the case of $^4$He 
   including SRC, Eq.~(\ref{Eec2}) 
   (continuous line), 
   and in the harmonic oscillator model, Eq.~(\ref{En2H}) 
   (dashed line). 
} 
\label{f4} 
\end{figure} 
\begin{figure} 
\centerline{\epsfig{figure=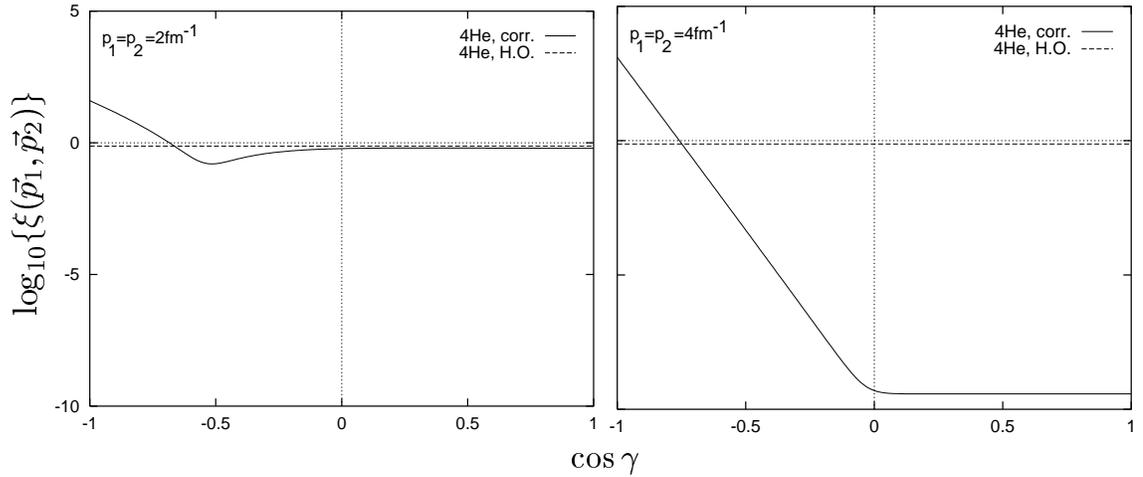,width=15cm}} 
\caption{ 
   The quantity $\xi (\vc{p}_1,\vc{p}_2)$ 
   (see Eq.~(\ref{Exi})) 
   in logarithmic scale, 
   as a function of $\cos{\gamma}$ 
   ($\gamma$ is the angle between the vectors $\vc{p}_1$ and $\vc{p}_2$) 
   for $p_1=p_2=$~2,~4~fm$^{-1}$, 
   in the case of $^4$He 
   including SRC 
   (continuous line), 
   and in the harmonic oscillator model 
   (dashed line). 
} 
\label{f5} 
\end{figure} 

\end{document}